\documentclass[aps,prb,twocolumn,groupedaddress,showpacs,showkeys]{revtex4-1}
\usepackage{graphics}
\usepackage{amsmath}
\usepackage{amssymb}
\usepackage{color}
\newcommand{\bb}{\mathbf}
\newcommand{\nn}{\mathrm}

\begin{document}
\title{Unusual boundary effect on coherency of	 two-band superconductivity}
\author{Artjom Vargunin}
\affiliation{Institute of Physics, University of Tartu, Tartu, EE-50411, Estonia, \\ 
Department of Theoretical Physics, The Royal Institute of Technology, Stockholm, SE-10691 Sweden}
\author{K\"ullike R\"ago}
\affiliation{Institute of Physics, University of Tartu, Tartu, EE-50411, Estonia}
\author{Teet \"Ord}
\affiliation{Institute of Physics, University of Tartu, Tartu, EE-50411, Estonia}

\date{\today}

\begin{abstract}
We demonstrate that the healing of two-band superconductivity near the surface of the system is governed by length scales which are drastically different from correlation lengths. Similar to the one-band case, one of the characteristic lengths diverges at critical temperature $T_\nn{c}$, while another one shows unusual behaviour having singularity at $T_{\nn{c}+}<T_\nn{c}$. By moving away from the boundary, these scales approach coherence lengths in the bulk state, where the divergence at $T_{\nn{c}+}$ is removed by arbitrary weak interband coupling.
Such a boundary-driven modification in coherency affects proximity phenomenon. We show that critical temperature of binary NS system can exhibit an inflection near $T_{\nn{c}+}$ as two-band superconducting layer becomes thinner. 
\end{abstract}

\pacs{}
\keywords{}

\maketitle

\section{Introduction}
Correlation length is one of the most fundamental characteristics for superconducting state. According to the single-gap theory, this length arises in almost all problems of inhomogeneous superconductivity, including surfaces, interfaces, defects, and vortices.

Spatial coherency in multigap superconductors had attracted much interest recently due to prediction of unconventional magnetic response called type-1.5 regime\cite{babaev3}. The phenomenon manifests itself as vortex clustering due to multi-scale physics resulting in the short-range type-II-like intervortex repulsion and long-range type-I-like attraction. Vortex distributions observed in multiband systems MgB$_2$\cite{moshchalkov0}, Sr$_2$RuO$_4$\cite{ray} and pnictides\cite{vinnikov} support type-1.5 scenario.

The specific features of the coherency peculiar for two-band superconductivity were first 
discussed almost thirty years ago by means of two-component Ginzburg-Landau equations\cite{poluektov}. 
It was shown that density variations are determined by the mixture of coherency modes characterized by qualitatively different length scales: conventional one which diverges at phase transition temperature and unconventional one which stays finite. 
Microscopic calculations fully confirm the accuracy of that scheme \cite{babaev1}. At that, disparity between correlation lengths can be significant as was demonstrated for different types of intercomponent interaction involving Josephson\cite{vargunin3}, mixed gradient\cite{babaev4}, or density-density couplings\cite{babaev5}. Such a disparity results in the qualitatively different density variations in the bands allowing unconventional magnetic response. 

In this contribution, we address coherency properties of two-band superconductor near its boundary, where system is affected by proximity effect as was pointed by Meissner\cite{meissner}. 
At present, proximity effects in hybrid structures involving multiband superconductors are under active investigation. Proximity can shed light on multiband nature of superconducting layer\cite{brinkman} and pairing symmetry\cite{spm}. The Josephson junctions with two-band superconductors revealed the significance of time-reversal-symmetry-breaking \cite{ng} for multiband superconducting state, for instance, in iron-based compounds \cite{stanev1}. This feature can result in the exotic behaviour such as negative proximity effect when $s$-wave superconductivity is suppressed by contact with nominally stronger superconductor\cite{stanev2}.

In what follows we examine the universality of coherency channels by moving from the bulk region to the boundary of a two-band $s$-wave superconductor. The proximity effect, i.e. mutual influence of the subsystems in contact, 
is taken into account phenomenologically by using Ginzburg-Landau formalism with relevant boundary conditions at the interface\cite{degennes1}. Qualitative estimates by means of McMillan tunneling model\cite{mcmillan} or Green's functions approach\cite{usadel} are out of the scope of present research. We found that there is non-trivial boundary driven modifications in coherency which cannot be anticipated from single-band superconductivity model.

\section{Two-component Ginzburg-Landau equation and its asymptotics}
We start with the Ginzburg-Landau formalism for clean two-band ($\alpha=1,2$) superconductor in zero magnetic field. By taking into account Josephson tunneling between bands, the free-energy density $F=\sum_\alpha F_\alpha$ is given by
\begin{align}
F_\alpha=a_\alpha|\delta_\alpha|^2+b_\alpha|\delta_\alpha|^4/2-c\delta_\alpha\delta_{3-\alpha}^\ast+K_\alpha|\nabla\delta_\alpha|^2,
\end{align}
where expansion coefficients read as \cite{dao}
\begin{align}
&a_\alpha=W_{3-\alpha,3-\alpha}/W^2-\rho_\alpha\ln\left[0.13\hbar\omega_\nn{D}/(
k_\nn{B}T)\right],\nonumber\\
&b_\alpha=0.11\rho_\alpha/(k_\nn{B}T)^2,\qquad c=W_{12}/W^2,\\
& K_\alpha=b_\alpha\hbar^2v_{\nn{F}\alpha}^2/6,\qquad W^2=W_{11}W_{22}-W_{12}^2,\nonumber
\end{align}
Here $W_{\alpha\alpha}>0$ and $W_{12}=W_{21}$ are matrix elements for intraband and interband pair-transfer interaction channels, $\rho_\alpha$ is the density of states at the Fermi level, and $v_{\mathrm{F}\alpha}$ is the Fermi
velocity in the corresponding band. Electron-electron interactions are assumed to be non-zero and independent on electron wave
vector in the Debye layer $\pm\hbar\omega_\mathrm{D}$ around chemical potential.

Spatial variation of gap fields $\delta_\alpha$ is governed by the two-component Ginzburg-Landau equation
\begin{align}\label{GLeqs}
a_\alpha\delta_\alpha-c\delta_{3-\alpha}+b_\alpha|\delta_\alpha|^2\delta_\alpha=K_\alpha \nabla^2\delta_\alpha,
\end{align}
supplemented by relevant boundary condition. Following de Gennes\cite{degennes}, we require
$\mathbf{n}\nabla\delta_\alpha=\sum_\beta\tau_{\alpha\beta}\delta_\alpha$ at the boundary of a superconductor, where $\bb{n}$ is unit vector normal to the surface and $\tau_{\alpha\beta}$ is matrix of inverse extrapolation lengths whose coefficients satisfy $\tau_{\alpha\alpha}=\tau_{\alpha\alpha}^\ast$ and $\tau_{21}=\tau_{12}^\ast K_1/K_2$. To preserve time-reversal symmetry on a level of boundary condition, we require an invariance under transformation $\delta_\alpha\mapsto\delta_\alpha^\ast$, which results in $\tau_{12}=\tau_{12}^\ast$.

Extrapolation lengths entering matrix $\hat \tau$ characterize the interaction between superconductor and ambient environment and, in particular, the penetration of order parameters into it. For example, for vacuum or insulators, extrapolation lengths approach infinity, since Cooper pairs cannot penetrate into these materials. For normal metal, extrapolation lengths are finite so that superconductivity is nucleated near the surface of the metal. In magnetic materials, spins of charged carriers are aligned in the same direction. Since in the $s$-wave superconductor the spins of electrons in a Cooper pair are antiparallel, neighbourship with ferromagnetic metal strongly suppresses superconductivity and extrapolation length is vanishing. Above boundary conditions can be used also for contact between different superconductors, if Josephson current is absent. In this case the states with broken time-reversal symmetry can appear near interface \cite{ng}. Moreover, one can expect the surface enhancement of superconductivity for the border with other superconductor which has higher critical temperature. In that case extrapolation length is negative. Note also that extrapolation length can be adjusted by manipulating abrasive grain size, annealing parameters\cite{kozhevnikov}, or external electric field\cite{lipavsky1}.

Before we analyse various regimes of Ginzburg-Landau equation (\ref{GLeqs}) we rescale quantities to have dimensionless units. By introducing length $\lambda=\Phi/\sqrt{8\pi\sum_\alpha K_\alpha\Delta_\alpha^2}$, where $\Phi=\hbar c/(2e)$ and $\Delta_\alpha$ are gap values in the bulk, we
scale distances as $X=\lambda x$ and order parameters as $\delta_\alpha=\Delta_\alpha f_\alpha e^{i\psi_\alpha}$, where it is assumed that gap fields depend only on a distance $x$ to the interface. In this case Ginzburg-Landau equation (\ref{GLeqs}) can be written by means of amplitudes $f_\alpha$ and phase difference $\psi=\psi_1-\psi_2$ in the form
\begin{align}\label{GLeqsp}
&\mathcal{A}_\alpha f_\alpha+\mathcal{C}_\alpha^2G^2\psi^{\prime2}/f_\alpha^3-\mathcal{C}_\alpha f_{3-\alpha}\cos\psi+\mathcal{B}_\alpha f_\alpha^3=f_\alpha^{\prime\prime},\nonumber\\
&f_1f_2\sin\psi=(G\psi^\prime)^\prime,\qquad \psi_2^\prime=-\mathcal{C}_2G\psi^\prime/f_2^2,
\end{align}
where $G=f_1^2f_2^2/(\mathcal{C}_2 f_1^2+\mathcal{C}_1 f_2^2)$ and $\mathcal{A}_\alpha=a_\alpha\lambda^2/K_\alpha$, $\mathcal{B}_\alpha=b_\alpha\Delta_\alpha^2\lambda^2/K_\alpha$ and $\mathcal{C}_\alpha=c\Delta_{3-\alpha}\lambda^2/(K_\alpha\Delta_\alpha)$. Spatial dependence of the phases generates supercurrent in the bands, however, net supercurrent is vanishing.

In homogeneous case one has $f_\alpha=1$ and $\psi=0$ assuming that interband coupling constant is positive so that Eq. (\ref{GLeqsp}) defines unambiguously bulk gaps $\Delta_\alpha$. The single-band limit of Eq. (\ref{GLeqsp}) corresponds to the equation for stronger-superconductivity component, where $\mathcal{C}_\alpha=0$ and $-\mathcal{A}_\alpha=\mathcal{B}_\alpha\equiv\kappa^2$ should be taken. Note that the parametrization of a non-degenerate two-band system requires more than one parameter $\kappa$.

\subsection{Behaviour near bulk}
Let us first analyse gap field behaviour near bulk state $f_\alpha=1$, where minimal free energy is achieved when $\psi=0$ for $W_{12}>0$. By introducing small deviations $\epsilon_\alpha=1-f_\alpha$, restoration of the gap-order parameters is given by
$(\mathcal{A}_\alpha+3\mathcal{B}_\alpha)\epsilon_\alpha-\mathcal{C}_\alpha\epsilon_{3-\alpha}=\epsilon_\alpha^{\prime\prime}$. To decouple these equations, we 
introduce mixing angles $\varphi_\pm$ and new fields $\epsilon_\pm$ as follows
\begin{align}
\epsilon_1=\cos\varphi_-\epsilon_--\sin\varphi_+\epsilon_+,\\
\epsilon_2=\sin\varphi_-\epsilon_-+\cos\varphi_+\epsilon_+.\nonumber
\end{align}
Choice $\tan\varphi_-=\mathcal{C}_2\tan\varphi_+/\mathcal{C}_1=2(\xi_-^2-\xi_1^2)/(\mathcal{C}_1\xi_1^2\xi_-^2)$, where $2\xi_\alpha^{-2}=\mathcal{A}_\alpha+3\mathcal{B}_\alpha$,
decouples equations as $\epsilon_\pm^{\prime\prime}=2\epsilon_\pm/\xi_\pm^2$. The letter defines correlation lengths $\xi_\pm$ for a two-band system as
\begin{align}\label{xipm}
2\xi_\pm^{-2}=\xi_1^{-2}+\xi_2^{-2}\pm\sqrt{(\xi_1^{-2}-\xi_2^{-2})^2+\mathcal{C}_1\mathcal{C}_2}.
\end{align}
Note that correlation lengths $\xi_\pm$ are dimensionless and given in the units of $\lambda$.

In the absence of interband coupling, each condensate has its own coherency channel characterized by dimensionless length scale $\xi_\alpha$. Interband pairing mixes these channels and restoration of gap-order parameters towards their bulk values in the joint condensate consists of two modes scaled by $\xi_\pm$. Mixing has two profound effects on coherency. First, it removes critical point due to superconducting instability of the weaker-superconductivity component because the joint condensate has single critical temperature. Second, the temperature dependence of dimensional correlation lengths $\xi_\pm\lambda$ becomes very dissimilar: $\xi_-\lambda$ behaves critically diverging at the phase-transition point $T_\nn{c}$, while $\xi_+\lambda$ stays always finite\cite{babaev1,vargunin1}.

Note that the bulk state is characterized by additional length due to interband phase variation $(\mathcal{C}_1+\mathcal{C}_2)\psi=\psi^{\prime\prime}$. Here $\mathcal{C}_1+\mathcal{C}_2$ is dimensionless mass of the Leggett mode.

\subsection{Behaviour near boundary}
Near-boundary state can have nontrivial interband phase which deviates from its bulk value. By using Taylor series anzats $f_\alpha=\sum_{n=0}^\infty f_\alpha^{(n)} x^n/n!$, analogously for $\psi$, one can construct solution starting from $f_\alpha^{(0)}$ and $\psi^{(0)}$ given.
Boundary conditions leads to $f_\alpha^{(1)}=\mathcal{T}_\alpha f_\alpha^{(0)}+\mathcal{C}_\alpha \mathcal{T}f_{3-\alpha}^{(0)}\cos\psi^{(0)}$ and $\psi^{(1)}=-\mathcal{T}f_1^{(0)}f_2^{(0)}\sin\psi^{(0)}/G^{(0)}$, where $\mathcal{T}_\alpha=\lambda\tau_{\alpha\alpha}$ and $\mathcal{T}=K_1\tau_{12}/(c\lambda)$. Next-order coefficients follows from Ginzburg-Landau equation (\ref{GLeqsp}).

Let us find an asymptotic near the interface of a superconductor. We assume that interface suppresses superconductivity as for the border with magnetic material or normal metal, and neglect time-reversal symmetry breakdown $\psi^{(0)}\mod{\pi}=0$. In this case, the Ginzburg-Landau equations can be linearized $\mathcal{A}_\alpha f_\alpha\mp\mathcal{C}_\alpha f_{3-\alpha}=f_\alpha^{\prime\prime}$, where upper sign corresponds to $\psi^{(0)}=0$. We decouple bands using mixed modes
\begin{align}\label{fpm}
f_1=\cos\theta_-f_--\sin\theta_+f_+\\
f_2=\sin\theta_-f_-+\cos\theta_+f_+.\nonumber
\end{align}
By taking $\tan\theta_-=\mathcal{C}_2\tan\theta_+/\mathcal{C}_1=\pm2(\Gamma_--\Gamma_1)/(\mathcal{C}_1\Gamma_-\Gamma_1)$, where $\Gamma_\alpha=2/\mathcal{A}_\alpha$, we obtain $f_\pm^{\prime\prime}=2f_\pm/\Gamma_\pm$ 
with
\begin{align}
2/\Gamma_\pm=1/\Gamma_1+1/\Gamma_2\pm\sqrt{(1/\Gamma_1-1/\Gamma_2)^2+\mathcal{C}_1\mathcal{C}_2}.
\end{align}
Dimensionless length scales of the problem are given by $\gamma_\pm=\sqrt{|\Gamma_\pm|}$. 
The latter ones define the restoration of two-band superconductivity close to its interface. One easily finds that dimensional lengths $\gamma_\pm\lambda$ diverge when $\mathcal{A}_1\mathcal{A}_2=\mathcal{C}_1\mathcal{C}_2$ which is satisfied at the temperatures $T_{\mathrm{c}\pm}$, respectively, 
with
\begin{align}
2/\ln\left[1.13\hbar\omega_\mathrm{D}/(k_\mathrm{B}T_{\nn{c}\pm})\right]=\sigma
\mp\sqrt{\sigma^2-4\rho_1\rho_2W^2},
\end{align}
where $\sigma=\sum_\alpha\rho_\alpha W_{\alpha\alpha}$. The larger one of these temperatures is the critical temperature of the joint condensate $T_\mathrm{c}=T_{\nn{c}-}>T_{\nn{c}+}$. Another one has physical meaning only for
vanishing interband interaction when $T_{\nn{c}-}\to T_{\nn{c}1}$ and $T_{\nn{c}+}\to T_{\nn{c}2}$, where $T_{\mathrm{c}\alpha}$ 
are critical temperatures for non-interacting condensates.

If the near-boundary state breaks time-reversal symmetry, the linearized equation can be approximately written in the form $\mathcal{\tilde A}_\alpha f_\alpha-\mathcal{\tilde C}_\alpha f_{3-\alpha}=f_\alpha^{\prime\prime}$, where it is assumed that interband phase varies weakly and $\mathcal{\tilde A}_\alpha=\mathcal{A}_\alpha+\mathcal{C}_\alpha^2 \mathcal{T}^2f_{3-\alpha}^{(0)2}\sin^2\psi^{(0)}/f_{\alpha}^{(0)2}$ and $\mathcal{\tilde C}_\alpha=\mathcal{C}_\alpha\cos\psi^{(0)}$. In this case characteristic lengths have more involved temperature behaviour.

\section{Results and discussion}
\subsection{Length scales}
\begin{figure}[t]
\begin{center}
\resizebox{0.98\columnwidth}{!}{\includegraphics{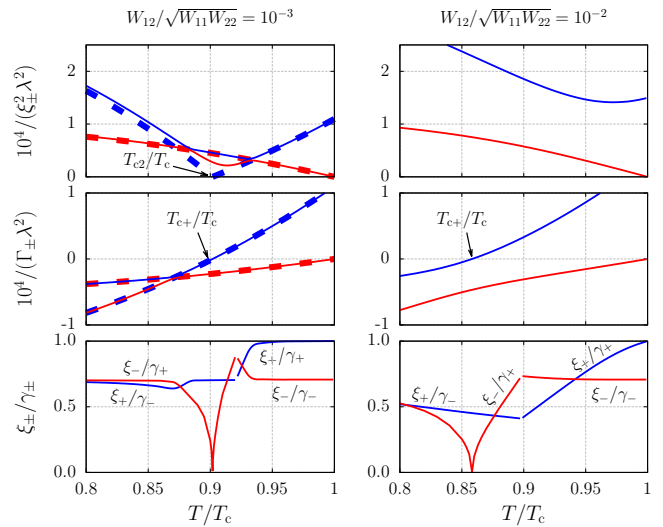}}
\end{center}
\vspace{-0.6cm}\caption{The temperature behaviour of characteristic masses for weak (left column) and intermediate (right column) interband couplings. Upper and middle rows of panels represent dependencies for inverse values of $(\xi_-\lambda)^2$, $\Gamma_-\lambda^2$ (red) and $(\xi_+\lambda)^2$, $\Gamma_+\lambda^2$ (blue curves). For weak interband interaction, the dependence of one-band counterparts $1/(\xi_\alpha\lambda)^2$ and $1/(\Gamma_\alpha\lambda^2)$ is also depicted by dashed curves. By taking $T_{\nn{c}2}=0.9T_{\nn{c}1}$ for non-interacting condensates, one observes the vanishing of $1/(\xi_{1,2}\lambda)^2$ and $1/(\Gamma_{1,2}\lambda^2)$ at corresponding critical temperatures $T_{\nn{c}1,2}$. In the lower row of panels the fractions $\mathrm{max}(\xi_\pm)/\mathrm{max}(\gamma_\pm)$ and $\mathrm{min}(\xi_\pm)/\mathrm{min}(\gamma_\pm)$ are shown by red and blue colours, correspondingly. The parameters are $\rho_{1,2}=(1;0.97)\ 1/(\nn{eV}\cdot\nn{cell})$, $v_{\nn{F}1,2}=(10;5)\times10^5\ \nn{m}/\nn{s}$ and $W_{\alpha\alpha}=0.3\ \nn{eV}\cdot\nn{cell}$. }\label{f1}
\end{figure}

The conventional superconductivity teaches us that coherency properties are always scaled by single length called correlation length. This scale is responsible for spatial evolution of the order parameter near interface as well as at the bulk of the system. In a two-band case we have more involved situation, since two-band superconductor possesses two characteristic length scales in the bulk state and two scales near the boundary. Let us now follow the formation of these lengths under interband interaction and interrelations between them.

The hybridization of former autonomous lengths is taking place as follows. Interband pairing results in one diverging ($\xi_-\lambda$) and one finite ($\xi_+\lambda$) correlation length instead of former ones $\xi_\alpha\lambda$ which are diverging at different temperatures $T_{\nn{c}\alpha}$, see Fig. \ref{f1}. Hybridization leads to the smearing of critical behaviour at the phase transition point of the weaker-superconductivity component and avoided crossing points appear. As a result, $\xi_->\xi_+$ for all temperatures.

The hybridization of $\Gamma_\alpha$ under interband pairing occurs in slightly different manner. The intersection is transformed into avoided crossing point, see Fig. \ref{f1}, however, the vanishing of $1/\Gamma_+$ at $T_{\nn{c}+}$ remains even for finite interband interaction. Therefore, discontinuity in the behaviour of $\Gamma_2$ at $T_{\nn{c}2}$ is not removed by mixing of the condensates and it is always seen in the behaviour of $\Gamma_+$ at $T_{\nn{c}+}$. As a result, $\gamma_->\gamma_+$ only in the vicinity of $T_\nn{c}$, while deeper in the superconducting phase $\gamma_-<\gamma_+$.	

In a conventional superconductor, the proportionality between lengths operating in various domains of spatially confined system is fixed, $\xi/\gamma=1/\sqrt{2}$. One can say that there is a single scale responsible for spatial evolution of the order parameter near interface as well as at the bulk of the system. In two-band situation, the bulk state has two characteristic lengths $\xi_\pm$. By approaching an interface from the bulk side, both the longer and shorter scales lengthens, but this happens very ununiformly. The ratios $\mathrm{max}(\xi_\pm)/\mathrm{max}(\gamma_\pm)$ and $\mathrm{min}(\xi_+,\xi_-)/\mathrm{min}(\gamma_\pm)$ are very sensitive to the temperature and interband coupling, see Fig. \ref{f1}. At the critical point, the ratios reach values expected, $\xi_-/\gamma_-=1/\sqrt{2}$ and $\xi_+/\gamma_+=1$. However, such an universality of characteristic length-scales holds only in finite temperature interval close to $T_\nn{c}$ which drastically shrinks under interband coupling.

In contrast to single-gap systems, we obtained that generally there is no such a length which scales the healing of two-band superconductivity near interface and in the bulk simultaneously. The letter is consequence of more basic fact that correlation length is a characteristic of bulk state only so that it would be not necessarily present near interface. In particular, it explains why singularity of $\gamma_+$ at $T_{\mathrm{c}+}$ is not a signature of any phase transition taking place in the system.

Qualitative difference between characteristic lengths at the bulk and near the interface of a two-band superconductor can be demonstrated analytically. Recently it was shown \cite{shanenko,vargunin2} that effect of interband interaction can be understood in terms of external field which affects the superconducting instability of the weaker-superconductivity component. Such a field acts in a same fashion as an external electric field on ferroelectric phase transition. As a result, dimensional correlation length is scaled at $T_{\mathrm{c}2}$ as $\xi_-\lambda\propto W_{12}^{-1/3}$ for weak interband coupling. Here, critical index $1/3$ is in full agreement with Landau theory of criticality meaning that interband interaction smears criticality in the weaker superconductivity component as an external field does with the Landau phase transition. At the same time, we obtain $\gamma_+\lambda\propto W_{12}^{-1}$ at $T_{\nn{c}2}$. Very different exponent suggests that scales found near boundary cannot be treated as real correlation lengths of a two-band superconductor.

\subsection{Spatial coherency modes for gap-order parameters}
\begin{figure}[t]
\begin{center}
\resizebox{0.98\columnwidth}{!}{\includegraphics{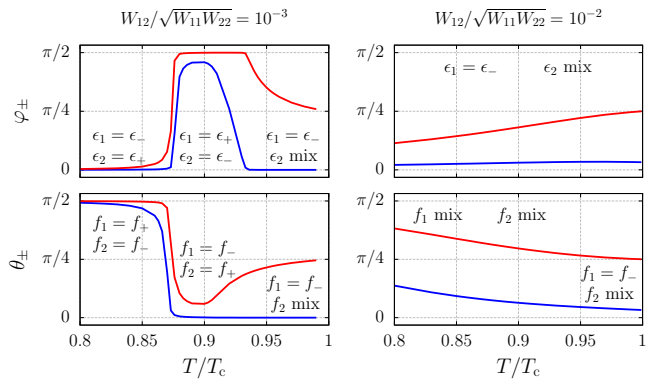}}
\end{center}
\vspace{-0.6cm}\caption{The temperature behaviour of mixing angles $\varphi_-$, $\theta_-$ (red) and $\varphi_+$, $\theta_+$ (blue curves) for weak (left column) and intermediate (right column) interband couplings. For different temperature intervals, the dominant mode for band order parameters is specified, where abbreviation "mix" means that both modes are present.}\label{f2}
\end{figure}

To get full understanding of the boundary effects on coherency, we discuss coherency modes of two-band superconductor in details. Here we have multiple modes scaled by different characteristic lengths. The modes contributes to the inhomogeneous gap functions with certain weight given by mixing angles, $\varphi_\pm$ in the bulk and $\theta_\pm$ near boundary. Let us follow the distribution of coherency modes for each band.

First, we discuss weak interband interaction. In this case there is an alternation of the modes contributing band order parameters as temperature changes, see Fig. \ref{f2}. In the vicinity of critical temperature, $\varphi_+$ and $\theta_+$ are small which means that spatial evolution of the stronger-superconductivity gap is dominated by the mode with longer (critical) length scale, $\xi_-$ and $\gamma_-$, correspondingly. At the same time weaker-superconductivity gap demonstrates strong mode mixing.

As temperature lowers, the modes start switching-over in the band channels due to rapid change of the mixing angles. The latter takes place at avoided crossing points of characteristic lengths, cf Figs. \ref{f1}, \ref{f2}. In the bulk region, mode contribution is changed for the first time near former instability of the weaker-superconductivity component, $T_{\nn{c}2}$. In this domain, $\epsilon_1$ consists primarily of the mode with shorter length, while $\epsilon_2$ of the mode with longer length. Interestingly, similar picture holds near interface as well, not due to the mode switching but because $\gamma_+$ becomes the longest of the lengths and $\theta_-$ is suppressed, see Fig. \ref{f2}. As a result, $\delta_1$ modifies near boundary rapidly, while $\delta_2$ changes with distance very slowly, similar to the gap evolution in the bulk state. By decreasing the temperature further, one 
passes avoided crossing point so that  dominance of the modes in the band channels becomes again inverted: mode with longer length dominates for stronger, while mode with shorter length for weaker-superconductivity gap.

We observe that for weak interband coupling the composition of the band coherency channel is not 
affected by the boundary effects: if band gap changes in space slowly (rapidly) in the bulk, it would vary slowly (rapidly) near interface as well so that rapidly (slowly) oscillating mode does not admix. There are only quantitative changes in the characteristic length due to the interface. Stronger interband interaction mixes modes more effectively, especially near interface, see Fig. \ref{f2}.

\begin{figure}[t]
\begin{center}
\resizebox{0.98\columnwidth}{!}{\includegraphics{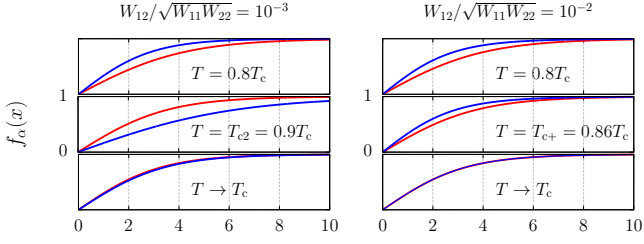}}
\end{center}
\vspace{-0.6cm}\caption{Spatial restoration of $f_1$ (red) and $f_2$ (blue) for weaker (left) and stronger (right column) interband coupling. 
}\label{f3}
\end{figure}

The picture provided by the analysis of mixing angles can be easily recognized in the spatial evolution of gap-order parameters. To show that, we solve inhomogeneous Ginzburg-Landau equations (\ref{GLeqsp}) numerically by taking $f_\alpha(x=0)=0$ and $\psi=0$. If interband interaction is weak, stronger-superconductivity gap changes with distance slower than weaker-superconductivity order parameter at lower temperatures, however, the picture becomes inverted in the vicinity of $T_{\nn{c}2}$, see Fig. \ref{f3}. That behaviour is caused by the alternation of the modes with longer/shorter characteristic length in the band gap as temperature increases. For stronger interband coupling such an alternation of the modes is not observable due to effective mixing. For the same reason, slow restoration of the mode with characteristic length $\gamma_+$ does not show up in the spatial behaviour of gap-order parameters at $T_{\nn{c}+}$.

The striking feature seen in Fig. \ref{f3} is identical spatial behaviour of the gap fields in the limit $T\to T_\mathrm{c}$. This peculiarity is general for two-band superconductivity and can be explained as follows. The Ginzburg-Landau functional contains Josephson-like coupling between gap-order parameters, however one can introduce new fields so that Josephson tunneling between them is eliminated at the cost of multiple fourth-order interactions. It appears \cite{babaev2} that one of new fields does not survive in the limit $T\to T_\mathrm{c}$ so that theory reduces to the single-component model.

\begin{figure}[t]
\begin{center}
\resizebox{0.98\columnwidth}{!}{\includegraphics{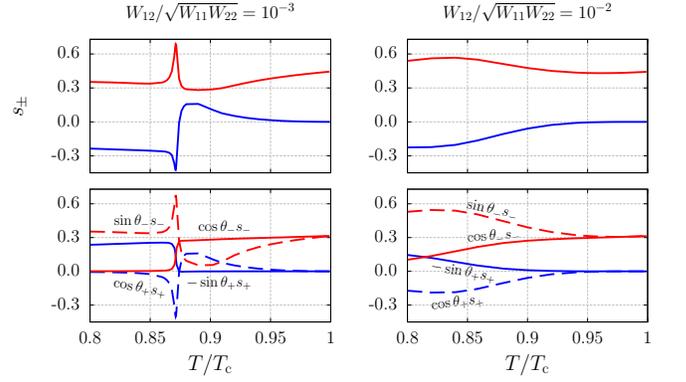}}
\end{center}
\vspace{-0.6cm}\caption{The temperature behaviour of slope $s_-$ (red) and $s_+$ (blue) for weaker (left column) and stronger (right column) interband coupling. In the lower row of panels we demonstrate how modes weighted by mixing angles contribute to the stronger-superconductivity gap (solid) and weaker-superconductivity gap (dashed) close to the boundary.}\label{f4}
\end{figure}

To check this conclusion, we calculate the temperature driven contribution of coherency modes into gap fields. Near interface of the system, the asymptotics are given by $f_\pm^{\prime\prime}=2f_\pm/\Gamma_\pm$, where $\Gamma_+$ changes sign at $T_{\mathrm{c}+}$, therefore two temperature domains should be analysed separately. In the region $T_{\nn{c}+}<T<T_\nn{c}$ we take $\Gamma_-=-\gamma_-^2$ and $\Gamma_+=\gamma_+^2$, so that solutions read as $f_-=\nu_-\sin(\sqrt{2}x/\gamma_-)$ and $f_+=\nu_+\sinh(\sqrt{2}x/\gamma_+)$. In the region $T<T_{\nn{c}+}$ we obtain $\Gamma_\pm=-\gamma_\pm^2$ and $f_\pm=\nu_\pm\sin(\sqrt{2}x/\gamma_\pm)$. The amplitudes $\nu_\pm$ are defined by initial slopes for the gaps $f_\alpha^\prime(0)$ and in both regions
\begin{align}\label{nupm}
&\sqrt{2}\nu_-\cos(\theta_--\theta_+)=\gamma_-[\sin\theta_+f_2^\prime(0)+\cos\theta_+f_1^\prime(0)],\nonumber\\
&\sqrt{2}\nu_+\cos(\theta_--\theta_+)=\gamma_+[\cos\theta_-f_2^\prime(0)-\sin\theta_-f_1^\prime(0)].
\end{align}
Close to the boundary, the asymptotics $f_\pm=s_\pm x$ are characterized by slopes $s_\pm=\sqrt{2}\nu_\pm/\gamma_\pm$. By using numerics we find $\nu_\pm$ and depict the temperature dependence of the slopes $s_\pm$ in Fig. \ref{f4}. We see that not only mixing angles but also amplitudes of the modes are of importance, especially close to $T_\nn{c}$, where non-critical mode drops out precisely due to peculiarities of the amplitudes. Namely, from definition (\ref{nupm}) we obtain $\nu_+=0$, if fields have identical slopes $f_1^\prime(0)=f_2^\prime(0)$ and   $\theta_-\to\pi/4$, which is exactly the case near $T_\nn{c}$. Note that the smaller interband coupling is, the closer to $T_\nn{c}$ one should be to obtain single-component model, see Fig. \ref{f4}.

\subsection{Proximity effects in a two-band superconductor}
The leakage of Cooper pairs from a superconductor is known for a long time \cite{Deutscher}. The property is being used for practical purposes due to ability to tune critical temperature, gap and shape of temperature dependencies via proximity effect \cite{gildemeister,kain,lacquaniti}.

Let us analyse how the multiple coherency modes near the interface of superconductor affect proximity phenomenon. We consider NS binary system, where the interface of two-band superconducting layer of thickness $L=\ell\lambda$ is covered by normal metal. Near the phase transition temperature $T_\nn{cNS}$ of the NS system the order parameters are small so that the Ginzburg-Landau equations have a form $\mathcal{A}_\alpha f_\alpha-\mathcal{C}_\alpha f_{3-\alpha}=f_\alpha^{\prime\prime}$ discussed above.
By introducing uncoupled modes $f_\pm$ be means of Eq. (\ref{fpm}), we obtain $f_\pm^{\prime\prime}=2f_\pm/\Gamma_\pm$ supplemented by the boundary conditions $f_\pm^\prime(\ell)=0$ and $f_\alpha^\prime(0)=\mathcal{T}_\alpha f_\alpha(0)$. The former is stemming from to the fact that S layer borders with vacuum at $x=\ell$ so that $f_\alpha^\prime(\ell)=0$, and the latter is a condition at NS interface $x=0$ where interband effect on the interpolation lengths is omitted.

The problem appears to be sensitive to the temperature domain. At lower temperatures $T_\nn{cNS}<T_{\nn{c}+}$, where $\Gamma_\pm=-\gamma_\pm^2$, the solution reads as $f_\pm(x)=\nu_\pm\cos[\sqrt{2}(x-\ell)/\gamma_\pm]/\cos(\sqrt{2}\ell/\gamma_\pm)$, and $\nu_\pm$ should be found from boundary condition at interface
\begin{align}\label{tcsn}
\nu_-\eta_{1-}\cos\theta_--\nu_+\eta_{1+}\sin\theta_+=0,\nonumber\\
\nu_-\eta_{2-}\sin\theta_-+\nu_+\eta_{2+}\cos\theta_+=0.
\end{align}
Here $\eta_{\alpha\pm}=\mathcal{T}_\alpha-\sqrt{2}\tan(\sqrt{2}\ell/\gamma_\pm)/\gamma_\pm$. The condition for non-trivial solutions $\nu_\pm\neq0$ defines the critical temperature $T_\nn{cNS}$ as function of $\ell$.

\begin{figure}[t]
\begin{center}
\resizebox{0.98\columnwidth}{!}{\includegraphics{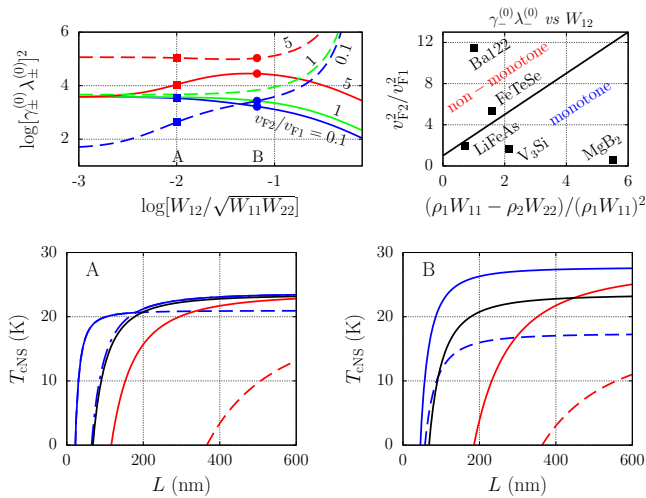}}
\end{center}
\vspace{-0.6cm}\caption{Top left: The behaviour of the slopes $\gamma_\pm^{(0)}\lambda_\pm^{(0)}$ (dashed/solid) as interband interaction increases for different values of $v_{\nn{F}2}$ indicated by number near each curve. For the two values of interband pairing marked by A and B we calculated the slopes shown by squares and circles, respectively. The evolution of $T_{\nn{cNS}}$ is illustrated in the lower row of panels for cases A and B, when $v_{\nn{F}2}/v_{\nn{F}1}=0.1;5$ (blue and red solid curves).  In these panels dashed curve is contribution from Eq. (\ref{TcNS+}), while dash-dot is Eq. (\ref{TcNS}). Note that critical temperature $T_\nn{cNS}$ is given by maximal of temperatures defined by these Eqs. Black curve
represents the one-band limit. Top right: Two regions in the space of ratios $v_{\nn{F}2}^2/v_{\nn{F}2}^2$ and $(\rho_1W_{11}-\rho_2W_{22})/(\rho_1W_{11})^2$ where $\gamma_-^{(0)}\lambda_-^{(0)}$ behaves qualitatively different under $W_{12}$: monotonically reduces or non-monotonically decreases passing through the maximum. Both these behaviours are present in the top left panel. Some relevant materials are also located using microscopic parameters from literature \cite{gamma_MgB2_V3Si,gamma_LiFeAs,gamma_FeTeSe,gamma_BaFeNiAs}. 
}\label{f5}
\end{figure}

The higher-temperature region $T_{\nn{c}+}<T_\nn{cNS}<T_\nn{c}$ differs from the previous case by $\Gamma_+=\gamma_+^2$ so that $f_+(x)=\nu_{+}\cosh[\sqrt{2}(x-\ell)/\gamma_+]/\cosh(\sqrt{2}\ell/\gamma_+)$, while $f_-(x)$ remains the same. Boundary condition at interface leads to Eq. (\ref{tcsn}), where substitution  $\eta_{\alpha+}=\mathcal{T}_\alpha+\sqrt{2}\tanh(\sqrt{2}\ell/\gamma_+)/\gamma_+$ should be made. 

The coherence length of the normal film at N side of proximity system describes the scale of penetration for the superconductivity.
Let us simplify the picture by assuming that the extrapolation lengths for both bands are close $\mathcal{T}_1\approx \mathcal{T}_2$ so that $\eta_{1\pm}\approx\eta_{2\pm}$. As a result, condition for $T_\nn{cNS}$ factorizes $\eta_{1+}\eta_{1-}=0$. When $T_\nn{cNS}>T_\nn{c+}$, only $\eta_{1-}$ can be vanishing which leads to implicit expression for critical temperature $\ell+1/\mathcal{T}_1=\pi\gamma_-/(2\sqrt{2})$. Extrapolation length has regular temperature dependence. By expanding the Ginzburg-Landau coefficients in the vicinity of $T_\mathrm{c}=T_\mathrm{c-}$, where $\gamma_-\lambda$ is divergent, we obtain explicit form
\begin{align}\label{TcNS}
&T_\nn{cNS}/T_\nn{c}=1-1.23\left[\gamma_-^{(0)}\lambda_-^{(0)}\right]^2/L^2,
\end{align}
where $L\gg1/\tau_{11}$ is taken. When $T_\nn{cNS}<T_\nn{c+}$, the condition $\eta_{1+}\eta_{1-}=0$ can have two solutions with largest one given by $\eta_{1+}=0$. In this case we expand the Ginzburg-Landau coefficients near $T_{\nn{c}+}$, where the singular behaviour of $\gamma_+$ leads to
\begin{align}\label{TcNS+}
&T_\nn{cNS}/T_{\nn{c}+}=1-1.23\left[\gamma_+^{(0)}\lambda_+^{(0)}\right]^2/L^2,\\
&\left[\gamma_\pm^{(0)}\lambda_\pm^{(0)}\right]^2=2(K_{2\pm}a_{1\pm}+K_{1\pm}a_{2\pm})/(\rho_2a_{1\pm}+\rho_1a_{2\pm}).\nonumber
\end{align}
Here $\gamma_\pm^{(0)}\lambda_\pm^{(0)}$ is dimensional length attributed to $\gamma_\pm$ in the leading order in $1-T/T_{\nn{c}\pm}$, respectively, when  
$T=0$ is substituted.

Fig. \ref{f5} (top left) demonstrates the slopes in dependencies (\ref{TcNS}) and (\ref{TcNS+}) for different interband couplings and Fermi velocities. The main feature at strong interband couplings is suppression of $\gamma_-^{(0)}\lambda_-^{(0)}$ and enhancement of $\gamma_+^{(0)}\lambda_+^{(0)}$ relative to the values $v_{\nn{F}1}/T_{\nn{c}1}$ and $v_{\nn{F}2}/T_{\nn{c}2}$, respectively, obtained for non-interacting bands. This means that critical temperature of NS sandwich decreases with $L$ slower than in binary system with one-band superconductor. The mode with $\gamma_+$ does not affect this process. Here the impact of interband interaction on the phase transition of NS system is equivalent to the bulk superconducting state, where interband pairing always enhances phase transition temperature of the condensate. 

The case with moderate and weak interband couplings is more interesting. The values of the slopes for non-interacting bands, $v_{\nn{F}\alpha}/T_{\nn{c}\alpha}$, points to possibility to manipulate the dependencies (\ref{TcNS}) and (\ref{TcNS+}) by band disparity. For weak interband interaction the situation is possible when $\gamma_-^{(0)}\lambda_-^{(0)}\gg\gamma_+^{(0)}\lambda_+^{(0)}$ provided that $v_{\nn{F}2}/v_{\nn{F}1}$ is small, see case A in top left panel in Fig. \ref{f5}. This means that the dependencies (\ref{TcNS}) and (\ref{TcNS+}) intersect so that the critical temperature $T_\nn{cNS}$ inflects close to $T_{\nn{c}+}$ when driving role switches between $\gamma_-$ and $\gamma_+$. Behaviour is shown in panel A in Fig. \ref{f5}.

In Fig. \ref{f5} (top right panel) we placed some multiband superconductors on the diagram according to the microscopic parameters taken from the fit to $\gamma$-model \cite{gamma_MgB2_V3Si}. The lower part of the diagram includes the region with small value $v_{\nn{F}2}/v_{\nn{F}1}$ where inflection of $T_\nn{cNS}(L)$ is possible provided that interband interaction is sufficiently weak. Unfortunately, materials listed in the diagram do not meet this requirement.  

Non-monotonic behaviour of $T_\nn{cNS}$ can be an explanation for seeming linear suppression of transition temperature by film thickness found in \cite{doi}. Inflection of $T_\nn{cNS}(L)$ seen in panel A in Fig. \ref{f5} straighten the critical temperature of NS system being considered as the function of reciprocal value of superconducting layer thickness. According to Eqs. (\ref{TcNS})-(\ref{TcNS+}), in this case $T_\nn{cNS}(1/L)$ represents an envelope consisting of intersecting concave parabolas. One of these parabolas has higher vertex located at $T_\nn{c}$ and short focal length defined by $[\gamma_-^{(0)}\lambda_-^{(0)}]^{-2}$. Second parabola has lower vertex at $T_{\nn{c}+}$ and longer focal length defined by $[\gamma_+^{(0)}\lambda_+^{(0)}]^{-2}$. The resulting envelope can be approximated by linear dependence much better than single concave parabola appearing in the case of conventional superconductivity.

Moderate interband pairing can strongly enhance the slope $\gamma_-^{(0)}\lambda_-^{(0)}$ above its one-band limit $v_{\nn{F}1}/T_{\nn{c}1}$. This regime becomes visible when $v_{\nn{F}2}/v_{\nn{F}1}$ is large so that $\gamma_-^{(0)}\lambda_-^{(0)}$ passes through the maximum as a function of $W_{12}$, see case B in top left panel in Fig. \ref{f5}. In this case the critical temperature of NS system with two-band superconducting layer becomes strongly suppressed by finite layer thickness so that $T_\nn{cNS}$ appears to be smaller than its single-band counterpart, see panel B in Fig. \ref{f5}. Interestingly, this is opposed to the bulk superconductivity, where interband interaction always enhances phase transition temperature of the condensate. For layered superconductor, the coupling with additional superconductivity component can suppress the critical temperature. The behaviour is peculiar for materials from the upper part of the diagram given in Fig. \ref{f5}, possibly for Ba$122$ compound.

The results discussed in this section can be of relevance for multiband superconducting films. These systems may be affected by the presence of degraded non-superconducting layers at the film surface or film-substrate interface resulting in the suppression of superconductivity due to the proximity effect\cite{zhang}.

%
%
%
%
%

\section{Conclusion}
Spatial evolution of conventional superconductivity is governed by single correlation length. In a two-band superconductor, the coherency can be decomposed by two independent channels, each one characterized by its own length scale. We analysed evolution of these channels as one moves from bulk state to the boundary. Although qualitative features of these channels remain invariant, interesting quantitative modifications appear. Namely, in the bulk two-band superconductor, the restoration of superconductivity is driven by two correlation lengths: one is finite and another one divergent at $T_\nn{c}$. However, near an interface two singular length scales which diverge at different temperatures, $T_\nn{c}$ and $T_{\nn{c}+}$, respectively, are present. The existence of additional characteristic length which becomes infinitely large at $T_{\nn{c}+}<T_\nn{c}$ can manifest itself in the proximity phenomena, for instance, in the inflection of transition temperature of NS binary system as two-band superconducting layer becomes thinner. Complex character of coherency at the interface of a two-band superconductor can lead to the interband-coupling driven suppression of critical temperature of NS system compared to its single-band counterpart. This behaviour is opposite to the bulk state where interband interaction always enhances the critical temperature of the condensate.

\section*{Acknowledgement}
The study was supported by the Estonian Ministry of Education and Research through the Institutional Research Funding IUT2-27 and through the grant PUTJD141.

\bibliography{bbib}
\end{document}